\documentclass[12pt]{iopart}   
\usepackage{graphicx}
\usepackage{epsfig}
\begin{document}

\title[A non-conserving coagulation model with extremal dynamics]{A non-conserving coagulation model with extremal dynamics}

\author{R\'obert Juh\'asz} 
\address{Research Institute for Solid
State Physics and Optics, H-1525 Budapest, P.O.Box 49, Hungary}
\ead{juhasz@szfki.hu} 

\begin{abstract}
A coagulation process is studied in a set of random masses, in
which two randomly chosen masses and the smallest mass of the set multiplied
by some fixed parameter $\omega\in [-1,1]$ are iteratively added. 
Besides masses (or primary variables), secondary variables are also considered
that are correlated with primary variables and coagulate according to the above
rule with $\omega=0$. 
This process interpolates between known statistical physical models: 
The case $\omega=-1$ corresponds to the strong disorder renormalisation group
transformation of certain disordered quantum spin chains whereas 
$\omega=1$ describes coarsening in the one-dimensional Glauber-Ising model. 
The case $\omega=0$ is related to the renormalisation group transformation of
a recently introduced graph with a fat-tail edge-length distribution. 
In the intermediate range $-1<\omega<1$, the exponents  $\alpha_{\omega}$ and $\beta_{\omega}$ that characterise the growth
of the primary and secondary variable, respectively, are accurately estimated 
by analysing the differential equations describing the process in the
continuum formulation. 
According to the results, the exponent $\alpha_{\omega}$ varies
monotonically with $\omega$ while $\beta_{\omega}$ has a maximum at $\omega=0$.

\end{abstract}

\maketitle

\newcommand{\bc}{\begin{center}}
\newcommand{\ec}{\end{center}}
\newcommand{\be}{\begin{equation}}
\newcommand{\ee}{\end{equation}}
\newcommand{\beqn}{\begin{eqnarray}}
\newcommand{\eeqn}{\end{eqnarray}}

\section{Introduction}

Coagulation processes arise in various areas of physics; one may think of
polymerisation, growth of ordered domains in non-equilibrium magnetic systems 
\cite{bray}, dynamics of droplets when water condenses on non-wetting 
surfaces \cite{dgy}, etc. 
The substance, or ``mass'' that aggregates is very frequently not conserved
during the process: for example, 
agglomerating insoluble inclusions in molten metal
may be lost from the melt by attachment to the wall of the vessel \cite{wmg}.
Therefore the theoretical investigation of the kinetics of such non-conserving
coagulation processes is of great importance.  
Moreover, the models developed for the description of such systems may 
show interesting behaviour: the Smoluchowsky equation with certain coagulation
kernels exhibits gelation transition and, in general, even the simplest
models with conserved mass may have non-trivial solutions, see
e.g. Ref. \cite{wmg} and references therein. 
Beside quite realistic ones, 
there is a special class of (possibly non-conserving) coagulation models  
where only the actually smallest one among the masses is active 
while the other masses are temporarily inert.   
This type of {\it extremal dynamics} can be regarded as a rough 
approximation for models where the reaction rates are 
decreasing functions of the mass of particles.  
In what follows, we shall survey three processes with extremal
dynamics in detail. 
We mention, however, that models of this type have also been 
introduced in the
context of dynamics of growing and coalescing droplets \cite{dgy} or
multispecies pair annihilation reactions \cite{dht}.  

In the one-dimensional Glauber-Ising model started from a random initial state
at zero temperature, the domain walls move as independent random walkers and
annihilate upon meeting. While the closest pairs of walls come together and
annihilate, the other domain walls hardly move. A simplified model of
evolution of distances $X_i$ between adjacent walls can be formulated as
follows \cite{nagai,kon}. The shortest interval $X_m$ is eliminated
together with the two adjacent intervals $X_1$ and $X_2$ and replaced by 
$\tilde X=X_1+X_2+X_m$. 
As the density of walls tends to zero,
the distributions of intervals at different times become self-similar,
depending on a single time-dependent length scale, and the corresponding
scaling function can be calculated exactly \cite{nagai,bdg,rutenberg}.  
Another quantity of interest is the fraction of space which has never been
traversed by a domain wall. The length $Y_i$ of such parts of intervals
transforms in the way $\tilde Y=Y_1+Y_2$ when the shortest interval is
eliminated. 
The characteristic value of $X$ depends on
the fraction $c$ of the initial intervals that have not yet been eliminated 
as $X\sim c^{-\alpha}$, obviously, with $\alpha=1$, while 
it has been found that $Y\sim c^{-\beta}$, where the persistence 
exponent $\beta=0.824924 12\dots$ is the zero of a 
transcendental equation \cite{bdg}.   
In addition to this, the autocorrelation exponent has also been exactly
calculated in this model \cite{bd}. To obtain this quantity, the overlap $Z_i$
of an interval with its initial state that transforms as $\tilde
Z=Z_1+Z_2-Z_m$ had to be considered.
Later, a generalisation of persistence has been studied in the
same model, which required to introduce an auxiliary variable transforming 
as $\tilde Y=Y_1+Y_2+pY_m$ \cite{mb}. Here, 
the generalised persistence exponent has
been found to vary monotonically with the partial survival factor $p$
in the range $-1\le p\le 1$.

The next example is the strong disorder renormalisation group transformation 
of inhomogeneous quantum spin chains \cite{mdh}. Here, the degrees of freedom
related to the largest coupling (a bond between neighbouring spins or a local
external field) are eliminated one after the other. 
In terms of logarithmic couplings, $X_i$, the
renormalisation rule generally reads as $\tilde X=X_1+X_2-X_m$, 
where $\tilde X$ is a newly formed effective variable and $X_1$,$X_2$ 
are variables adjacent to the smallest one, $X_m$.
For the relation between these variables and the couplings in the particular 
 Hamiltonians we refer the reader to Ref. \cite{fisher}.   
A variable $Y_i$ that transforms according to  the rule 
$\tilde Y=Y_1+Y_2$ under such a
renormalisation step can be interpreted in the case of a particular model, the
transverse field Ising chain, as the magnetic moment of a spin. 
For this process with i.i.d. random initial variables
$X_i$, which corresponds to critical spin chains, the distribution of $X$
flows again to a fixed point where it shows scaling behaviour. 
The characteristic value of $X$ increases in the course of the process 
as $X\sim c^{-\alpha}$ with 
$\alpha=1/2$, while the variable $Y$ grows as $Y\sim c^{-\beta}$ with
$\beta=(1+\sqrt{5})/4=0.809016\dots$ \cite{fisher}.         
Note that the coagulation rules in the above two models 
differ only in the sign of $X_m$, 
which leads to different exponents $\alpha$ and $\beta$.
 
Our third example is a random graph where three edges emanate from each
node, and which is built on a regular one-dimensional lattice by adding long
edges in the following way. To each edge of the one-dimensional
lattice that we call short edges, a random weight $X_i$ is assigned. 
Defining the length of a path as the sum
of weights of the edges it contains, the closest pair of nodes of degree 2
with respect to this metric is chosen and connected by an edge of unit
weight. This step is then iterated until all nodes become of degree 3 
\cite{juhasz}. 
For this graph, a renormalisation procedure can be formulated 
where loops are eliminated 
step by step in reversed order compared to the construction procedure. 
Formally, the short edge with the minimal weight $X_m$ is eliminated
together with the nodes it connects, as well as with the 
neighbouring short edges with 
weights $X_1$, $X_2$ and a new effective short edge is formed with a weight
calculated asymptotically as $\tilde X=X_1+X_2$.  
According to numerical results, 
the characteristic value of effective weights grows as 
$X\sim c^{-\alpha}$ with $\alpha=0.826(1)$ \cite{juhasz}.
This exponent characterises at the same time the 
diameter of finite graphs with $N$ nodes with respect 
to the above metric via $D(N)\sim N^{\alpha}$. 

As can be seen, these seemingly different problems can be treated in a common
framework and can be interpreted as coagulation
processes with extremal dynamics. 
In the first example, the total sum of the variables $X_i$ is
conserved while in the latter two cases it is not. 
We will study in this work a coagulation model 
controlled by a parameter $\omega$ that interpolates 
continuously between the first two models and incorporates the third one as a
special case, as well. 
We are interested in the exponents $\alpha_{\omega}$ and $\beta_{\omega}$
for intermediate values of the parameter $\omega$ and shall
provide accurate estimates for $\alpha_{\omega}$ that is obtained 
as the root of a 
transcendental equation while $\beta_{\omega}$ is accurately determined 
by the numerical analysis of a system of non-linear differential equations. 
We shall see that $\alpha_{\omega}$ varies monotonically between 
the corresponding values of
the two marginal models, while, 
unlike the generalised persistence exponent of the model with
partial survival mentioned above \cite{mb}, the exponent $\beta_{\omega}$
shows a maximum when $\omega$ is varied. 
As can be seen, the transformation rule of the variable 
$Y$ does not depend directly on the
parameter $\omega$ but it is influenced indirectly via the correlations 
emerging between $X$ and $Y$, the strength of which is controlled
by $\omega$. Therefore our results may contribute to the
understanding of the role of correlations in such models.       
Moreover, these investigations provide an accurate estimate for the diameter
exponent of the graph quoted above, for which we obtain $\alpha=0.82617561$ in
agreement with the previous numerical result.   

The rest of the paper is organised as follows. In Section \ref{model},
the model and its continuum description is introduced. 
In Sections \ref{asec} and \ref{bsec}, the way of approximative
determination of the exponents $\alpha_{\omega}$ and $\beta_{\omega}$
is presented. Some calculations are given in the Appendix. Finally,
results are discussed in Section \ref{disc}. 

\section{The model and its continuum formulation}
\label{model}

\subsection{Definition of the model}

Let us consider a finite set of positive 
vectors $V_i=(X_i,Y_i)$ indexed by the
integers $i=1,2,\dots,N$. We assume, moreover, that $N$ is odd. 
The vectors are independent, identically distributed random variables 
drawn from a continuous distribution $\rho(X,Y)dXdY$, for which we require 
that all moments exist.
The first components $X_i$ and the second components $Y_i$ are called
primary and secondary variables, respectively.  
Assume, furthermore, that $\omega\in [-1,1]$ is a fixed real number. 
Now, the following procedure is considered on this set. 
The vector $V_m$ with the smallest
primary variable is chosen and, at the same time, two further vectors $V_i$ and
$V_j$ are chosen at random from the set. These three vectors are removed
and a new vector $\tilde V$ with components
\beqn
\tilde X=X_i+X_j+\omega X_m \nonumber \\
\tilde Y=Y_i+Y_j 
\label{rules}
\eeqn
is added to the set.
Thereby the number of vectors in the set is reduced by
two. Note that the vectors remain independent after such an operation and
that 
\be
\tilde X\ge X_i,X_j,X_m
\label{ineq}
\ee
even for $\omega=-1$.   
This step is then iterated until a single vector $V_N=(X_N,Y_N)$ is left in
the set.
In this general formulation, the cases $\omega=1,-1,0$ correspond to
the three models in the order as they were quoted in the Introduction.  
Based on the known asymptotical behaviour of $X_N$ and $Y_N$ for 
large $N$ in the marginal cases $\omega=-1,1$, 
we expect
\be
X_N\sim N^{\alpha_{\omega}} \quad {\rm and} \quad Y_N\sim N^{\beta_{\omega}}
\label{powerlaw}
\ee 
to hold also for intermediate parameter values $-1<\omega<1$
with some exponents $\alpha_{\omega}$ and $\beta_{\omega}$ that may 
depend on $\omega$. 

\subsection{Continuum formulation}

Now, we consider the continuum limit $N\to\infty$ and introduce 
the probability density $P_{\Gamma}(X)$ of the primary variable 
that has the support 
$\Gamma\le X<\infty$ and that depends on the lower boundary 
$\Gamma$ as a parameter. 
The function $P_{\Gamma}(X)$ is normalised as $\int_{\Gamma}^{\infty}P_{\Gamma}(X)dX=1$ for any $\Gamma$.
Following Ref. \cite{bdg}, we consider, furthermore, the expected value 
$\overline{Y}_{\Gamma}(X)$ of the secondary
variable under the condition that the primary variable is $X$.        
In the continuum limit, the system is described by these two
functions of $X$, which depend on the lower boundary of the support 
$\Gamma$ as a parameter.
The inequality (\ref{ineq}) implies that,
as the fraction of vectors $c_{\Gamma}$ that have not yet been eliminated decreases in
the course of the coagulation process, the lower edge $\Gamma$ 
of the distribution continuously increases.          
As it is shown in the Appendix, one may write 
the following differential equation for $P_{\Gamma}(X)$:
\be
\frac{\partial P_{\Gamma}(X)}{\partial \Gamma}=P_{\Gamma}(\Gamma)
\Theta[X-(2+\omega)\Gamma]\int_{\Gamma}^{X-(1+\omega)\Gamma}P_{\Gamma}(X')P_{\Gamma}(X-X'-\omega\Gamma)dX',
\label{Pdiff}
\ee
where $\Theta(X)$ is the Heaviside step function. 
The fraction $c_{\Gamma}$ 
is related to $\Gamma$ as 
$dc_{\Gamma}/c_{\Gamma}=-2P_{\Gamma}(\Gamma)d\Gamma$ or, equivalently,
\be 
\frac{dc_{\Gamma}}{d\Gamma}=-2P_{\Gamma}(\Gamma)c_{\Gamma}.
\label{cdiff}
\ee
The function $Q_{\Gamma}(X)$ defined as 
\be
Q_{\Gamma}(X)\equiv P_{\Gamma}(X)\overline{Y}_{\Gamma}(X),
\label{Qfunc}
\ee
can be shown to obey the differential equation  
\be
\frac{\partial Q_{\Gamma}(X)}{\partial \Gamma}=2P_{\Gamma}(\Gamma)\Theta[X-(2+\omega)\Gamma]\int_{\Gamma}^{X-(1+\omega)\Gamma}Q_{\Gamma}(X')P_{\Gamma}(X-X'-\omega\Gamma)dX'.
\label{Qdiff}
\ee 
The derivation of this equation is given again in the Appendix. 

\subsection{Fixed point solution}

In the marginal cases $\omega=-1,1$, it is known that, for any well-behaving
initial distributions $\rho(X,Y)$ with finite moments, the 
solutions of Eqs. (\ref{Pdiff}) and (\ref{Qdiff}) tend to a universal 
fixed point solution $P^*_{\Gamma}(X)$, $Q^*_{\Gamma}(Y)$ 
in the limit $\Gamma\to\infty$ that has the scaling property 
\beqn 
P^*_{\Gamma}(X)=\Gamma^{-1}f(X/\Gamma)  \nonumber \\
Q^*_{\Gamma}(X)=\Gamma^{\delta_{\omega}-1}g(X/\Gamma),
\label{fp} 
\eeqn
with some number $\delta_{\omega}$ that is related to the growth
exponents as\footnote{This can be seen from the equation 
$\overline{Y}^*_{\Gamma}(\Gamma)\equiv
Q^*_{\Gamma}(\Gamma)/P^*_{\Gamma}(\Gamma)=\Gamma^{\delta_{\omega}}g(1)/f(1)$
that indicates the asymptotical relation $Y\sim X^{\delta_{\omega}}$ between the
typical values of primary and secondary variables.} 
\be 
\delta_{\omega}=\beta_{\omega}/\alpha_{\omega}.
\label{delta}
\ee 
Therefore we expect this to hold also for intermediate parameter values  
$-1<\omega<1$ with some (a priori unknown) exponent $\delta_{\omega}$ 
that may depend on $\omega$. 
Indeed, the functions in Eq. (\ref{fp}) solve Eqs. (\ref{Pdiff}) and
(\ref{Qdiff}) provided that the universal 
scaling functions $f(x)$ and $g(x)$
satisfy the following differential equations:
\beqn
\frac{d[xf(x)]}{dx}=-f_1\Theta(x-2-\omega)\int_1^{x-1-\omega}f(x')f(x-x'-\omega)dx'
\label{fdiff}
\\
\frac{d[x^{1-\delta_{\omega}}g(x)]}{dx}x^{\delta_{\omega}}=-2f_1\Theta(x-2-\omega)\int_1^{x-1-\omega}g(x')f(x-x'-\omega)dx',
\label{gdiff}
\eeqn
where the notation $f_1\equiv f(1)$ has been used. 
For an alternative derivation of these equations in the case $\omega=1$, 
see Ref. \cite{bdg}.
Using the fixed point solution, Eq. (\ref{cdiff}) can be integrated yielding
the asymptotic relation in the large $\Gamma$ limit: 
\be 
\Gamma\sim c_{\Gamma}^{-\frac{1}{2f_1}}.
\ee 
Comparing this with Eq. (\ref{powerlaw}), we obtain the relation:
\be 
\alpha_{\omega}=\frac{1}{2f_1}.
\label{exprel}
\ee

\section{Approximative determination of $\alpha_{\omega}$} 
\label{asec}

As can be seen, Eq. (\ref{fdiff}) does not contain $g(x)$ and together
with Eq. (\ref{exprel}) it constitutes an autonomous problem for the calculation of the exponent $\alpha_{\omega}$. 
For the special case $\omega=-1$, the solution of Eq. (\ref{fdiff}) is of
simple form: $f(x)=e^{-x+1}$; this yields $\alpha_{-1}=\frac{1}{2}$. 
In the other marginal case, $\omega=1$, 
the Laplace transform of the solution is
known \cite{nagai,rutenberg} and $\alpha_{1}=1$. 
In the case $-1<\omega<1$, where Eq. (\ref{fdiff}) is not soluble, we shall
construct an approximative solution that enables us to give an 
accurate estimate
of $\alpha_{\omega}$. An alternative way related to the numerical analysis of
the Laplace transforms is presented in the next section. 

\begin{figure}[h]
\includegraphics[width=0.6\linewidth]{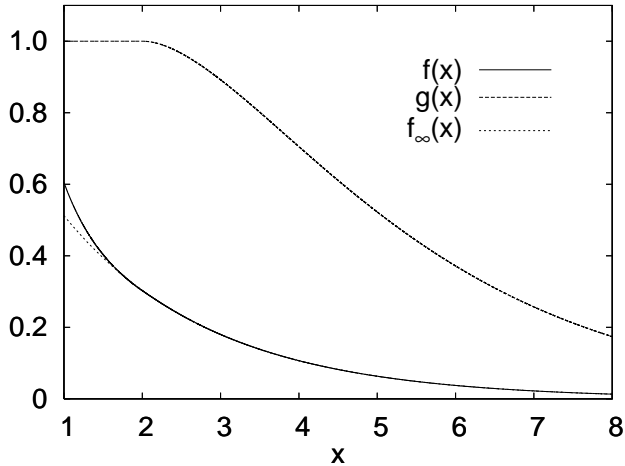}
\caption{\label{fig1} Graphs of the functions $f(x)$, $g(x)$ and $f_{\infty}(x)$ for $\omega=0$. The former two are obtained by
numerical integration of Eqs. (\ref{fdiff}) and (\ref{gdiff}), whereas the
latter is given in Eq. (\ref{asymp}).}  
\end{figure}

\begin{figure}[h]
\includegraphics[width=0.6\linewidth]{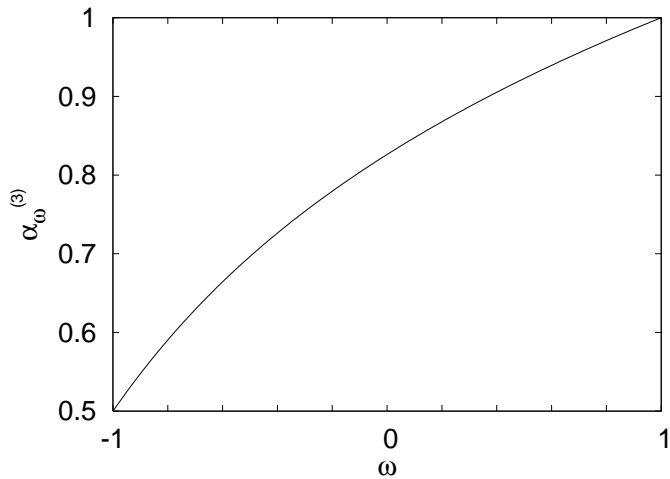}
\caption{\label{fig2} The third approximant $\alpha_{\omega}^{(3)}$ 
of the exponent $\alpha_{\omega}$ plotted against $\omega$.}
\end{figure}

Some properties of the scaling function $f(x)$ can be easily established
by investigating Eq. (\ref{fdiff}) without knowing the exact solution.
Apparently, the r.h.s. of Eq. (\ref{fdiff}) and, as a consequence, 
$f(x)$ is non-analytical at $x=x_1\equiv 2+\omega$. 
But, as $f(x)$ itself appears on the r.h.s. as a convolution with a shifted
argument $x-1-\omega$, the r.h.s. as well as $f(x)$ 
must be non-analytical also at 
$x=x_2\equiv x_1+1+\omega$. Iterating this argument, it turns out that there
are infinitely many points where $f(x)$ is non-analytical.
To be precise, one can show by recursion that the 
$2n$th derivative of $f(x)$ is discontinuous at\footnote{For a more direct way to this result in the case $\omega=1$, where the explicit
form of the scaling function $f(x)$ is available, see Ref. \cite{rutenberg}.} 
\be 
x_n=1+(1+\omega)n, \qquad n=0,1,2,\dots.
\ee 
Furthermore, the function value of $f(x)$ at some $x'$ is determined 
by $f(x)$ in the restricted domain $(1,x'-1-\omega)$. 
Due to this property, $f(x)$ can be constructed 
in the intervals $[x_n,x_{n+1}]$ step by step 
starting with $n=0$. However, the solution is more and more complicated for
increasing $n$ as it contains multiple integrals that 
cannot be evaluated analytically. 
In the domains $[x_n,x_{n+1}]$, $n=1,2,\dots$, the function $f(x)$ can be written in the following form:
\be 
f(x)=\frac{1}{x}\sum_{i=0}^nf_1^{2i+1}C^{(2i+1)}_{\omega}(x),
\quad  x_n\le x\le x_{n+1},   
\label{series}
\ee
whereas $f(x)=0$ if $x<x_0$.
The functions $C^{(2i+1)}_{\omega}(x)$ are independent of $f_1$ and the first
three of them read as 
\beqn 
C^{(1)}_{\omega}(x)=1, \\ \nonumber 
C^{(3)}_{\omega}(x)=-2\int_{x_1}^x\frac{\ln (x'-\omega
  -1)}{x'-\omega}dx', \\ \nonumber
C^{(5)}_{\omega}(x)=-2\int_{x_2}^x\int_{x_0}^{x'-x_2+1}\frac{C^{(3)}_{\omega}(x'-x''-\omega)}{x''(x'-x''-\omega)}dx''dx'.
\eeqn
Substituting an exponential trial function in Eq. (\ref{fdiff}), we obtain that
the asymptotical solution $f_{\infty}(x)$ in the limit $x\to\infty$ is 
of the form  
\be 
f_{\infty}(x)=\frac{a}{f_1}e^{-a(x+\omega)},
\label{asymp}
\ee
with the some number $a$ that is not fixed by this substitution. 
The graph of $f(x)$ for $\omega=0$ is shown in Fig.\ref{fig1}.
As can be seen, $f(x)$ tends rapidly to $f_{\infty}(x)$ for
increasing $x$. 
This suggests an approximation for $f(x)$ in which $f(x)$ is replaced by
 the simple asymptotical function $f_{\infty}(x)$ for large $x$. 
To be precise, the $n$th ($n=0,1,2,\dots$) approximant $f^{(n)}(x)$ is defined as 
\beqn
f^{(n)}(x)=f(x), \qquad {\rm if} \quad x\le x_n \nonumber \\
f^{(n)}(x)=f_{\infty}(x), \qquad {\rm otherwise}.
\eeqn      
The two unknown parameters $f_1$ and $a$ are determined by the requirements
that $f^{(n)}(x)$ is continuous at $x=x_n$, i.e. 
\be 
f(x_n)=f_{\infty}(x_n),
\ee
and that it is normalised as 
\be 
\int_{x_0}^{x_n}f(x)dx+\int_{x_n}^{\infty}f_{\infty}(x)dx=1.
\ee
Using the expression in Eq. (\ref{series}), straightforward calculations
result in 
that the $n$th approximant  $f_1^{(n)}$ ($n>0$) is the root of the
following transcendental equation: 
\beqn 
\sum_{i=0}^{n-1}[f_1^{(n)}]^{2i+1}C^{(2i+1)}_{\omega}(x_n)+ \nonumber \\
+\frac{x_n}{x_n+\omega}
\left[1-\sum_{i=0}^{n-1}[f_1^{(n)}]^{2i+1}N^{(2i+1)}_{\omega}(x_n)\right]
\ln\left[f_1^{(n)}-\sum_{i=0}^{n-1}[f_1^{(n)}]^{2i+2}N^{(2i+1)}_{\omega}(x_n)\right]=0,
\nonumber \\
\label{trans}
\eeqn
where the function $N_{\omega}^{(2i+1)}(x)$ has been introduced as
\be 
N_{\omega}^{(2i+1)}(x)\equiv\int_{x_i}^{x}\frac{C^{(2i+1)}_{\omega}(x')}{x'}dx'.\label{nint}
\ee
We have numerically calculated the root of Eq. (\ref{trans}) 
and the $n$th approximant $\alpha_{\omega}^{(n)}$ of $\alpha_{\omega}$ 
by using Eq. (\ref{exprel}) for $n=1,2,3$ and
for several values of $\omega$. This has necessitated 
the numerical evaluation of
the integrals in Eq. (\ref{nint}) for $n>1$. 
Results are shown in Fig. \ref{fig2} and some numerical values are given in
Table I. As can be seen, the approximants $\alpha_{\omega}^{(n)}$ converge
rapidly with increasing $n$ and they increase monotonically with $\omega$. 
The best estimate for the diameter exponent of the graph cited in the
Introduction is $\alpha_0^{(3)}=0.82617561$. 

\begin{table}[h]
\begin{center}
\begin{tabular}{|r||c|c|c|c|c|}
\hline $\omega$ & $\alpha_{\omega}^{(1)}$ &$\alpha_{\omega}^{(2)}$
&$\alpha_{\omega}^{(3)}$ & $\delta_{\omega}$ 
& $\beta_{\omega}=\delta_{\omega}\alpha_{\omega}^{(3)}$\\
\hline
\hline -0.9 & 0.54752760 & 0.54752815 & 0.54752815 & 1.48973578 & 0.81567227\\
\hline -0.8 & 0.59036797 & 0.59037862 & 0.59037860 & 1.38841226 & 0.81968889\\
\hline -0.7 & 0.62906729 & 0.62911723 & 0.62911708 & 1.30687751 & 0.82217896\\
\hline -0.6 & 0.66421085 & 0.66434418 & 0.66434376 & 1.23995279 & 0.82375490\\
\hline -0.5 & 0.69632781 & 0.69659263 & 0.69659189 & 1.18399594 & 0.82476197\\
\hline -0.4 & 0.72586754 & 0.72630756 & 0.72630667 & 1.13643762 & 0.82540221\\
\hline -0.3 & 0.75320237 & 0.75385266 & 0.75385199 & 1.09543864 & 0.82579860\\
\hline -0.2 & 0.77863838 & 0.77952410 & 0.77952421 & 1.05965756 & 0.82602872\\
\hline -0.1 & 0.80242716 & 0.80356405 & 0.80356560 & 1.02809664 & 0.82614309\\
\hline  0.0 & 0.82477635 & 0.82617193 & 0.82617561 & 0.99999999 & 0.82617561\\
\hline 0.1 &  0.84585830 & 0.84751339 & 0.84751989 & 0.97478484 & 0.82614953\\
\hline 0.2 &  0.86581708 & 0.86772721 & 0.86773715 & 0.95199472 & 0.82608118\\
\hline 0.3 &  0.88477397 & 0.88693068 & 0.88694457 & 0.93126697 & 0.82598219\\
\hline 0.4 &  0.90283179 & 0.90522369 & 0.90524198 & 0.91230963 & 0.82586098\\
\hline 0.5 &  0.92007834 & 0.92269199 & 0.92271501 & 0.89488491 & 0.82572374\\
\hline 0.6 &  0.93658910 & 0.93940970 & 0.93943769 & 0.87879701 & 0.82557503\\
\hline 0.7 &  0.95242937 & 0.95544130 & 0.95547441 & 0.86388324 & 0.82541834\\
\hline 0.8 &  0.96765597 & 0.97084323 & 0.97088154 & 0.85000684 & 0.82525595\\
\hline 0.9 &  0.98231868 & 0.98566518 & 0.98570869 & 0.83705175 & 0.82508918\\
\hline 1.0 &  0.99646128 & 0.99995110 & 0.99999976 & 0.82492447 & 0.82492427\\
\hline
\end{tabular}
\end{center}
\caption{\label{table1} Approximants of the exponents $\alpha_{\omega}$, 
$\delta_{\omega}$ and $\beta_{\omega}$ for different values of $\omega$.} 
\end{table}

\section{Numerical determination of $\beta_{\omega}$} 
\label{bsec}

Next, we turn to the determination of the exponent $\delta_{\omega}$ (and, at
the same time, $\beta_{\omega}$ through Eq. (\ref{delta})), which
requires the analysis of the full problem, i.e. the system of differential
equations (\ref{fdiff}) and (\ref{gdiff}). 
Prior to this, 
a few remarks concerning the scaling function $g(x)$ are in order. 
First, as a consequence of the definition in Eq. (\ref{Qfunc}), $g(x)$ apparently 
inherits the singularity properties of $f(x)$ discussed in the previous
section. Furthermore, it can be written in a form analogous to
Eq. (\ref{series}). In the domain $[x_0,x_1]$, it has a simple form:
\be 
g(x)=g(1)x^{\delta_{\omega}-1}, \qquad  x_0\le x\le x_1.
\label{power}
\ee
Second, the differential equation (\ref{gdiff}) gives the scaling function
$g(x)$ only up to a multiplicative constant. 
This non-universal constant depends
on the initial distribution $\rho(X,Y)$ and it is fixed in a 
non-trivial way by the original equations (\ref{Pdiff}) and 
(\ref{Qdiff}) that are valid for any $\Gamma$.  
Third, the equation (\ref{gdiff}) contains the a priori unknown 
parameter $\delta_{\omega}$ that must be fixed by physical
considerations about the solution that depends on $\delta_{\omega}$.
Namely, the physically acceptable solution must be nonnegative and must have
the only reasonable asymptotics allowed by Eq. (\ref{gdiff}): 
\be
g_{\infty}(x)\simeq const\cdot xe^{-ax},
\ee
where the number $a$ is the same as that appears in Eq. (\ref{asymp}). 
Numerical analysis of Eq. (\ref{gdiff}) shows that these requirements are
fulfilled only for a single value of the parameter $\delta_{\omega}$. 

Following Ref. \cite{bdg}, it is, however, simpler to analyse 
the Laplace transform of
the equations (\ref{fdiff}) and (\ref{gdiff}). 
Introducing the functions 
\be
\phi(p)=\int_1^{\infty}e^{-px}f(x)dx, \qquad 
\psi(p)=\int_1^{\infty}e^{-px}g(x)dx,
\label{laplace}
\ee
the equations (\ref{fdiff}) and (\ref{gdiff}) transform to  
\beqn
p\phi'(p)=f_1[e^{-\omega p}\phi^2(p)-e^{-p}], 
\label{phidiff} \\
p\psi'(p)=-\delta_{\omega}\psi(p) -g_1e^{-p} + 2f_1e^{-\omega p}\psi(p)\phi(p),
\label{psidiff}
\eeqn
where the prime denotes derivation by $p$ and $g_1\equiv g(1)$. 
These equations are not soluble in the parameter range $-1<\omega<1$ but
asymptotical expressions of the solution can be established. 
The functions $\phi(p)$ and $\psi(p)$ have the small-$p$ expansions:
\be 
\phi(p)=\sum_{n=0}^{\infty}a_np^n, \qquad 
\psi(p)=g_1\sum_{n=0}^{\infty}b_np^n. 
\label{psiseries}
\ee
Substituting these into Eqs. (\ref{phidiff}) and (\ref{psidiff}), we obtain  
that the expansion coefficients for $-1<\omega<1$ are given by 
$a_0=1$, $b_0=\frac{1}{2f_1-\delta_{\omega}}$ and by the following recursion
relations for $n>0$\footnote{These series expansions are also valid for $\omega=-1$ with 
$a_2=5/2$, and for $\omega=1$ with  $a_1=-2e^{\gamma}$ \cite{bdg}, where $\gamma$ is Euler's constant, given by $\gamma=-\int_0^{\infty}\ln t e^{-t}dt=0.577215\dots$.}: 
\beqn
a_n=\frac{\frac{(-1)^n}{n!}(\omega^n-1) + 
\sum_{0\le i,j,k<n;~ i+j+k=n}\frac{(-\omega)^i}{i!}a_ja_k}{\frac{n}{f_1}-2} \\
b_n=\frac{\frac{(-1)^n}{n!}\left(\frac{\omega^n}{2f_1-\delta_{\omega}}-\frac{1}{2f_1}\right)
  + \frac{1}{2f_1-\delta_{\omega}}a_n +\sum_{0\le i,j,k<n;~
    i+j+k=n}\frac{(-\omega)^i}{i!}a_jb_k}{\frac{n+\delta_{\omega}}{2f_1}-1}. 
\eeqn

Next, we discuss the large-$p$ behaviour of $\psi(p)$. 
The differential equation (\ref{psidiff}) can have two kinds of 
asymptotical solutions depending on the parameter $\delta_{\omega}$.
If the second term on the r.h.s. dominates, we obtain 
\be 
\psi'(p)\simeq -g_1\frac{e^{-p}}{p},
\label{a1}
\ee  
while, if the first term dominates, we obtain
\be 
\psi(p)\simeq const\cdot p^{-\delta_{\omega}}.
\label{a2}
\ee
On the other hand, it follows from Eq. (\ref{laplace}) that 
$\psi(p)$ must have the large-$p$ asymptotics:
$\psi(p)\simeq g_1e^{-p}/p[1+O(1/p)]$. Using that the function $g(x)$ is
explicitely known in the domain $1\le x\le 2+\omega$ (see 
Eq. (\ref{power})), we can obtain a more accurate asymptotical form 
for $1/p\ll 1+\omega$. Replacing $g(x)$ by the function in Eq. (\ref{power})
in the entire domain $x\ge 1$, the integral in Eq. (\ref{laplace}) can be
evaluated, yielding
\be 
\psi_{\infty}(p)= g_1e^{-p}\sum_{k=0}^{\infty}
\left(\begin{array}{c}
\delta_{\omega}-1 \\ 
k
\end{array}\right)
\frac{k!}{p^{k+1}}, \qquad p\gg \frac{1}{1+\omega}.
\label{asympfull}
\ee
This shows that the physically acceptable asymptotics is that given 
in Eq. (\ref{a1}) whereas that in Eq. (\ref{a2}) is non-physical. 
Numerical analysis of the differential equations (\ref{phidiff}) and
(\ref{psidiff}) with the correct value of $f_1$ 
shows the following behaviour of the solution when the parameter $\delta_{\omega}$ is varied: 
For small enough $\delta_{\omega}$, the function 
$\psi(p)$ is non-monotonous and tends to zero from below 
for increasing $p$ as given in
Eq. (\ref{a2}); for large enough $\delta_{\omega}$, the function $\psi(p)$  
decays monotonically to zero again with the asymptotics given in
Eq. (\ref{a2}). These parameter regimes are separated by a ``critical'' value
of $\delta_{\omega}$. At this value, the solution decays monotonically to zero
with the physically acceptable asymptotics given in Eq. (\ref{a1}).
  
The numerical estimation of $\delta_{\omega}$ is based on this scenario: 
The differential equations (\ref{phidiff}) and
(\ref{psidiff}) are integrated from $p=0$ to some large $p$ 
and the true value of
$\delta_{\omega}$ is selected by the condition that $\psi(p)$ has the correct
asymptotics. 
We have assumed here that $f_1$ is already at our disposal. This can be
obtained either by the approximative procedure described in the 
previous section or, analogous to the above method, from the condition that
$\phi(p)$ has the correct asymptotics given by Eq. (\ref{asympfull}) 
with $\delta_{\omega}=0$.           

Before presenting numerical results on $\delta_{\omega}$, we show that,  
in the case $\omega=0$, the assumption on the uniqueness of the value
$\delta_{\omega}$ that corresponds to the correct asymptotics 
implies that $\delta_0=1$.
For $\omega=0$, the primary and the secondary variables coagulate according to
the same rules, see Eq. (\ref{rules}). If these variables are initially
perfectly correlated, i.e. $X_i=bY_i$ with some common constant 
$b$ for all $i$,  it is obvious that $\alpha_0=\beta_0$. 
Nevertheless, this equality holds for general initial distributions, as well. 
Indeed, it is easy to check that for $\omega=0$, the function
\be 
\psi(p)=-\frac{g_1}{f_1}\phi'(p)
\label{psi0}
\ee
solves Eq. (\ref{psidiff}) provided that
$\delta_0=1$ and $\phi(p)$ is the solution of
Eq. (\ref{phidiff}). 
In terms of the scaling functions, equation (\ref{psi0}) reads as 
$g(x)=xf(x)\frac{g_1}{f_1}$. 
As the asymptotics of the solution in Eq. (\ref{psi0}) is 
physically acceptable, we conclude that 
\be 
\delta_0=1.
\ee  

\begin{figure}[h]
\includegraphics[width=0.6\linewidth]{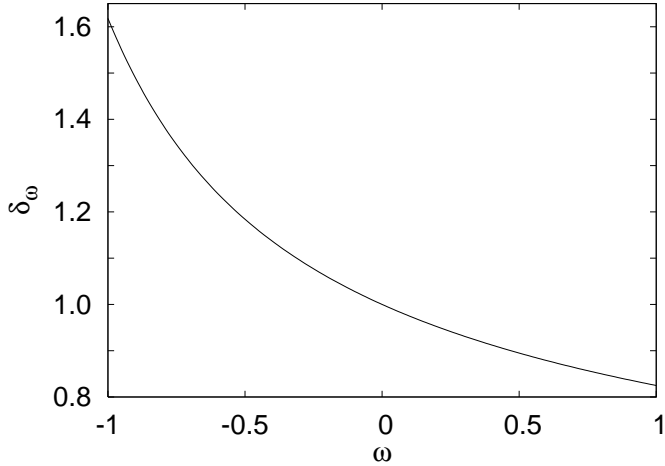}
\caption{\label{fig3} Numerically calculated exponent $\delta_{\omega}$ 
plotted against $\omega$.}
\end{figure}

\begin{figure}[h]
\includegraphics[width=0.6\linewidth]{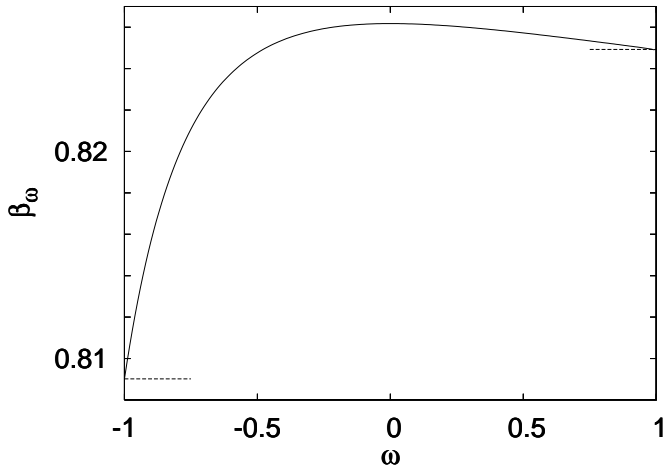}
\caption{\label{fig4} Numerically calculated exponent 
$\beta_{\omega}=\delta_{\omega}\alpha^{(3)}_{\omega}$ 
plotted against $\omega$. The horizontal lines indicate known values in the
marginal cases $\omega=-1,1$.}
\end{figure}

The details of the numerical determination of $\delta_{\omega}$ are the
followings. The best approximant that we have, $f_1^{(3)}$, has been
substituted in Eqs. (\ref{phidiff}) and (\ref{psidiff}) and a trial value 
for $\delta_{\omega}$ has been chosen.  
As the derivatives $\phi'(p)$ and $\psi'(p)$ calculated from these equations 
are of the form $0/0$ at $p=0$, the functions $\phi(p)$ and $\psi(p)$ 
were first calculated at $p=0.05$ by using the small-$p$ expansion in
Eq. (\ref{psiseries}) in order to avoid numerical uncertainties of the
integration in the vicinity of $p=0$. 
Then, starting from $p=0.05$, the differential equations were integrated by
the Bulirsch-Stoer method \cite{numrec} to some $p_f$, where $\psi(p_f)$ is
compared to the asymptotical form in Eq. (\ref{asympfull}). 
In practice, we have monitored the derivative $\psi'(p_f)$ rather than
$\psi(p_f)$ and $p_f=12$ was sufficiently large so that the asymptotical
value (at the critical $\delta_{\omega}$) is reached within the numerical
accuracy of the integration. 
The true $\delta_{\omega}$ was then
selected by the condition $\psi'(p_f)=\psi_{\infty}'(p_f)$. 
The exponent $\delta_{\omega}$ determined in this way is plotted against 
$\omega$ in Fig. \ref{fig3} whereas 
$\beta_{\omega}=\delta_{\omega}\alpha^{(3)}_{\omega}$ is plotted against
$\omega$ in Fig. \ref{fig4}.
Some numerical values can be found in Table \ref{table1}. 
As can be seen, $\delta_{\omega}$ decreases monotonically with $\omega$ but 
$\beta_{\omega}$ has a maximum at $\omega=0$. The latter observation will
be explained in the next section. 
 
\section{Discussion}
\label{disc}

We have shown that two problems, the renormalisation group procedure of certain
disordered quantum spin chains and a simple model describing coarsening in
the Glauber-Ising model can be deformed into each other by varying a single
parameter. 
The interpolating model is a special type of non-conserving 
coagulation process where only the actually smallest mass is active.   
In the range $-1\le\omega\le 1$, the exponents $\alpha_{\omega}$ and 
$\beta_{\omega}$ that characterise the
growth of the primary and the secondary variables, respectively,  
vary continuously with $\omega$. 
The latter exponent exhibits a maximum, which contrasts with the 
monotonous dependence of the generalised persistence exponent on 
the partial survival factor that appears directly 
in the transformation rule of the secondary variable \cite{mb}.  
Although, we have focused on the range $-1\le\omega\le 1$,
the equations written down in this work are valid also for $\omega >1$. 
In that case, the growth of the primary variable becomes super-linear, meaning
that $\alpha_{\omega}>1$. 

An intriguing feature of the process studied in this work is the universality
with respect to the initial distribution of the variables:
For a fixed $\omega$,
any sufficiently rapidly decaying initial distribution tends 
at late times to a
universal distribution that displays scaling.
Although, the process is universal in this sense, we have pointed out that
it is sensitive to the variations of the reaction rules parameterised by 
$\omega$. 
The dependence of $\alpha_{\omega}$ on $\omega$ is obvious since the
transformation rule of the primary variable contains $\omega$ explicitely.
The growth of the secondary variable is, however, 
affected by $\omega$ in a more subtle way.    
Focusing on the secondary variables, the difference to the process
of primary variables with $\omega=0$ 
is that, here, not exactly the smallest variable is
removed from the set. 
This is the reason for 
that $\beta_{\omega}$ is unequal to $\alpha_0$ for $\omega\neq 0$. 
Nevertheless, for any $\omega$, the removed secondary variable
is typically relatively small since $X_i$ and $Y_i$ become
positively correlated in the course of the process. 
Due to these correlations, the strength of which is controlled by 
$\omega$, the variation of $\beta_{\omega}$ is relatively slight. 
Indeed, it is by an order of magnitude
smaller than that of $\alpha_{\omega}$.

For $\omega=0$, we have shown that $\alpha_0=\beta_0$ even if the primary and
the secondary variables are initially not perfectly correlated. This can be
understood also on a microscopic level since, in this case,  
the vectors in the set are sums of an increasing number of 
initial vectors. Thus, the ratios $\tilde X_i/\tilde Y_i$ tend
stochastically to a common constant in the limit $\Gamma\to\infty$ 
for all $i$. In words, the two types of variables become asymptotically
perfectly correlated for $\omega=0$.      
Now, we are in a position to understand why the exponent $\beta_{\omega}$ 
is maximal at $\omega=0$.
At that point, the correlations are (at least asymptotically) perfect and
almost always the smallest one among the secondary variables is removed. 
For $\omega\neq 0$, however, the correlations are no longer perfect and, 
as a consequence, not strictly the smallest secondary variables are eliminated.
Therefore the fastest growth of $Y$ is realized at $\omega=0$.

In a general aspect, the benefit of the analysis carried out in this
work is that
the numerical technique developed here for obtaining accurate estimates of 
the growth exponents may also apply to other
non-soluble coagulation processes with extremal dynamics. 

\appendix
\section{}

When the primary variables in the infinitesimal interval 
($\Gamma$,$\Gamma+d\Gamma$) are eliminated in the course of the
process, we may write for the change of the probability 
density $P_{\Gamma}(X)$:
\beqn 
P_{\Gamma+d\Gamma}(X)=\left\{ P_{\Gamma}(X)+P_{\Gamma}(\Gamma)d\Gamma
\int dX_1\int dX_2P_{\Gamma}(X_1)P_{\Gamma}(X_2)\right.
\nonumber \\
\left.[ \delta(X-X_1-X_2-\omega \Gamma)-\delta(X-X_1)-\delta(X-X_2)]\right\}\frac{1}{1-2P_{\Gamma}(\Gamma)d\Gamma}.
\label{inf}
\eeqn 
Here, the first term of the integrand is related to the newly generated primary
variable while
the other two terms are related to the removed ones. 
The factor $1/[1-2P_{\Gamma}(\Gamma)d\Gamma]$ ensures that the distribution
remains normalised. 
Expanding the l.h.s. of Eq. (\ref{inf}) and retaining only terms of
order $d\Gamma$, we arrive at the differential equation (\ref{Pdiff}).

The expected value $\overline{\tilde Y}_{\Gamma}(X)$ 
of the newly generated secondary variable under the
condition that the generated primary variable is $X$ is given as
\beqn 
\overline{\tilde Y}_{\Gamma}(X)=\\ \nonumber
\left\{\int dX_1\int dX_2[\overline{Y}_{\Gamma}(X_1)+\overline{Y}_{\Gamma}(X_2)]P_{\Gamma}(X_1)P_{\Gamma}(X_2)\delta(X-X_1-X_2-\omega
\Gamma)\right\}/I_{\omega}(X)= \\ \nonumber
2\int_{\Gamma}^{X-(1+\omega)\Gamma}dX'P_{\Gamma}(X')\overline{Y}_{\Gamma}(X')P_{\Gamma}(X-X'-\omega\Gamma)/I_{\omega}(X),
\eeqn 
for $X>(2+\omega)\Gamma$, where the function
$I_{\omega}(X)\equiv\int_{\Gamma}^{X-(1+\omega)\Gamma}dX'P_{\Gamma}(X')P_{\Gamma}(X-X'-\omega\Gamma)$
has been introduced.
The expected value $\overline{Y}_{\Gamma+d\Gamma}(X)$ can then be written as 
the weighted average of $\overline{Y}_{\Gamma}(X)$ and
$\overline{\tilde Y}_{\Gamma}(X)$ as follows:
\beqn
\overline{Y}_{\Gamma+d\Gamma}(X)= 
\\ \nonumber
\frac{\left[P_{\Gamma}(X)-2d\Gamma
 P_{\Gamma}(\Gamma)P_{\Gamma}(X)\right] \overline{Y}_{\Gamma}(X)+
d\Gamma P_{\Gamma}(\Gamma)I_{\omega}(X)\overline{\tilde Y}_{\Gamma}(X)}
{P_{\Gamma}(X)+d\Gamma
 P_{\Gamma}(\Gamma)\left[I_{\omega}(X)-2P_{\Gamma}(X)\right]}.
\eeqn
This leads to the differential equation
\be 
\frac{\partial P_{\Gamma}(X)\overline{Y}_{\Gamma}(X)}{\partial \Gamma}=
P_{\Gamma}(\Gamma)\Theta[X-(2+\omega)\Gamma]I_{\omega}(X)\overline{\tilde Y}_{\Gamma}(X),
\label{Ydiff}
\ee
where we have made use of Eq. (\ref{Pdiff}).
Rewriting this equation in terms of $Q_{\Gamma}(X)$ given in Eq. (\ref{Qfunc}),
one arrives at Eq. (\ref{Qdiff}).

\ack
The author thanks F. Igl\'oi for useful discussions. 
This work has been supported by the Hungarian National Research Fund
under grant no. OTKA K75324.


\section*{References}
\begin{thebibliography}{99}

\bibitem{bray}
Bray A J, 1994 {\it Adv. Phys.} {\bf 43} 357

\bibitem{dgy} 
Derrida B, Godr\`eche C and Yekutieli I 1990 {\it Europhys. Lett.} {\bf 12}
385 

\nonum
Derrida B, Godr\`eche C and Yekutieli I 1991 {\it Phys. Rev. A} {\bf 44} 6241

\bibitem{wmg}
Wattis J A D, McCartney D G and Gudmundsson T 2004 {\it J. Eng. Math.} {\bf 49} 113  

\bibitem{dht}
Deloubri\`ere O, Hilhorst H J and T\"auber U C 2002 {\it Phys. Rev. Lett.}
{\bf 89} 250601 

\bibitem{nagai} 
Nagai T and Kawasaki K 1986 {\it Physica A} {\bf 134} 483

\bibitem{kon}
Kawasaki K, Ogawa A and Nagai T 1988 {\it Physica B} {\bf 149} 97 

\bibitem{bdg}
Bray A J, Derrida B and Godr\`eche C 1994 {\it Europhys. Lett.} {\bf 27} 175

\bibitem{rutenberg}
Rutenberg A D and Bray A J 1994 {\it Phys. Rev. E} {\bf 50} 1900 

\bibitem{bd} 
Bray A J and Derrida B 1995 {\it Phys. Rev. E} {\bf 51} R1633

\bibitem{mb}
Majumdar S N and Bray A J 1998 {\it Phys. Rev. Lett.} {\bf 81} 2626

\bibitem{mdh}
Ma S-K, Dasgupta C and Hu C-K 1979 {\it Phys. Rev. Lett.} {\bf 43} 1434

\nonum
For a review, see: Igl\'oi F and Monthus C 2005 {\it Phys. Rep.} {\bf 412} 277 

\bibitem{fisher}

Fisher D S 1992 {\it Phys. Rev. Lett.} {\bf 69} 534

\nonum 
Fisher D S 1994 {\it Phys. Rev. B} {\bf 50} 3799

\nonum
Fisher D S 1995 {\it Phys. Rev. B} {\bf 51} 6411

\bibitem{juhasz}
Juh\'asz R 2008 {\it Phys. Rev. E} {\bf 78} 066106

\bibitem{numrec}
Press W H, Teukolsky S A, Wetterling W T and Flannery B P 1992 {\it Numerical Recipes in C} Cambridge University Press, Cambridge

\end{thebibliography}
\end{document}